\documentclass[twocolumn,showpacs,amsmath,amssymb,floatfix]{revtex4}
\usepackage{graphicx}

\begin{document}

\title{Mesoscopic conductance effects in InMnAs structures}

\author{S. Lee$^{1}$, A. Trionfi$^{1}$, T. Schallenberg$^{2}$, H. Munekata$^{2}$, and D. Natelson$^{1}$}
\affiliation{$^{1}$Department of Physics and Astronomy, Rice University, Houston, TX 77005, USA}
\affiliation{$^{2}$Imaging Science and Engineering Laboratory, Tokyo Institute of Technology, 
Yokohama, Kanagawa 226-8503, Japan}

\date{\today}


\begin{abstract}
Quantum corrections to the electrical conduction of magnetic
semiconductors are comparatively unexplored.  We report measurements
of time-dependent universal conductance fluctuations (TDUCF) and
magnetic field dependent universal conductance fluctuations (MFUCF) in
micron-scale structures fabricated from two different
In$_{1-x}$Mn$_{x}$As thin films.  TDUCF and MFUCF increasing
in magnitude with decreasing temperature are observed.  At 4~K and
below, TDUCF are suppressed at finite magnetic fields independent
of field orientation.  
\end{abstract}

\maketitle

Mesoscopic phenomena such as weak localization (WL), Aharonov-Bohm 
(AB) oscillations, and universal conductance fluctuations (UCF) 
are comparatively unexplored in ferromagnetic (FM) systems.  
In part, this is because of complicating effects such as the 
anisotropic magnetoresistance (AMR) and the interplay between conduction
and FM domain structure.  
These materials are of fundamental interest since exchange 
correlations lead to collective degrees of freedom 
% ({\it e.g.} spin waves) 
not present in normal metals.  The interplay between 
FM order and coherence phenomena 
% ({\it e.g.} elastic vs. inelastic scattering of electrons by domain walls) 
is also a topic of interest\cite{HongetAl95PRB,TataraetAl97PRL}.  A
number of recent experiments have examined mesoscopic effects in FM
systems\cite{JaroszynskietAl95PRL,AprilietAl97SSC,AumentadoetAl00PB,
DugaevetAl01PRB,KasaietAl02PRL,KasaietAl03JAP,LeeetAl04PRB,WeietAl06PRL,
WagneretAl06PRL,VilaetAl06condmat}.  
Recent measurements\cite{KeizeretAl06Nature} in half-metallic
CrO$_{2}$ suggest that coherence lengths of hundreds of 
nanometers are possible in ferromagnetic systems.  The combination
of electronic coherence and magnetism may also enable 
device technologies.

Ferromagnetic semiconductors (FSs) are an interesting material
system to examine, since carrier-mediated spin 
exchange between Mn ions is thought to be the origin of the 
ferromagnetic phase\cite{DietletAl97PRB,KonigetAl00PRL}.  Two of 
the most studied FS systems\cite{Munekata06book} are Ga$_{1-x}$Mn$_x$As and its close 
relative In$_{1-x}$Mn$_x$As.  In these materials\cite{IyeetAl99MSEB}, 
one can achieve metallic conduction with ferromagnetic ordering at 
low temperatures with optimal doping, $x \sim 0.03-0.08$.

In this paper, we examine UCF in FM In$_{1-x}$Mn$_{x}$As thin films.
We perform both time-dependent UCF (TDUCF) and magnetic field-dependent
UCF (MFUCF) measurements in Hall bars patterned from two different wafers
at temperatures down to 2~K and magnetic fields, $B$, up to 9~T,
oriented both along the current direction and perpendicular to the
plane of the patterned FS film.  For $T >$~4~K, the TDUCF are nearly
independent of $B$.  At lower temperatures, the TDUCF noise power
decreases by roughly a factor of four as $B$ is raised above zero.
The magnitude and $B$ dependence of the TDUCF are independent of the
field orientation, suggesting that orbital effects are not the source
of the noise reduction.  The MFUCF data allow an order of magnitude
estimate of the coherence length at 2~K of $\sim$~50~nm.  The
magnitudes of the TDUCF and MFUCF are compared and discussed, as are
their temperature dependence and the nature of the low temperature
fluctuators that lead to the noise.

Provided films were grown on (100) GaAs wafers by molecular beam
epitaxy (MBE), starting with highly resistive 500~nm AlSb buffer
layers grown at a substrate temperature of 560~$^{\circ}$C with a
growth rate of 0.9~$\mu$m/h.  Then, 20~nm In$_{1-x}$Mn$_x$As films
were deposited at a substrate temperature of $\sim$ 200~$^\circ$C and
with a V/III flux ratio, $r = 3.4$.  Mn concentration was kept $<$
6~\% in order to avoid the formation of MnAs second phase.  Details of
the MBE process are reported in a recent separate
paper\cite{SchallenbergetAl06APL}.  Prepared in this manner, the
magnetic easy axis of the In$_{1-x}$Mn$_{x}$As is along the growth
direction.

Two different films were used to fabricate identically shaped wire
samples patterned by electron beam lithography.  A 1~keV Ar$^+$ ion
beam was used to sputter etch the extraneous film material.  AFM
measurements show that the widths of the wires and the narrow parts of
leads were $\sim$ 6.5~$\mu$m, and the center to center distances
between consecutive leads were 40~$\mu$m.  Table~\ref{tab:samples}
shows the specifications of the two samples.  The resistivities for
the processed material are several times higher than for the bulk films
(3-5~m$\Omega$-cm).  This suggests that the etching damages even
the ``unexposed'' semiconductor, though the magnetic properties
appear essentially unaffected by the etching.

\begin{table}
\caption{Parameters for both In$_{1-x}$Mn$_x$As samples used in the
experiments. Resistivity, carrier density, and mobility are calculated
from the sample resistance and the Hall resistance measured at 300~K,
3~T.  All samples have wires of width $\sim$ 6.5~$\mu$m and thickness
is 20~nm, and composed of six segments with length of 40~$\mu$m each.}
\begin{tabular}{c c c c c c c c c c c}
\hline \hline
Sample &~& $x$ &~& T$_{\rm{C}}$ [K] &~& $\rho$ [$\Omega~\rm{cm}$] 
&~& $p_{v}$ [$\rm{cm}^{-3}$] &~& $\mu$ [$\rm{cm}^{2} / \rm{V s}$] \\
\hline
\# 1 &~& 0.058 &~& 47 &~& 14.2$\times 10^{-3}$ &~& 1.87$\times 10^{20}$ &~& 2.34 \\
\# 2 &~& 0.045 &~& 27 &~& 11.2$\times 10^{-3}$ &~& 1.90$\times 10^{20}$ &~& 2.94 \\
\hline
\hline
\end{tabular}
\label{tab:samples}
\vspace{-3mm}
\end{table}

All measurements were performed in a $^4$He cryostat, with the
magnetic field either normal to the plane of the wire
(``perpendicular'') or along the wire axis (``parallel'').
Longitudinal and Hall resistances were measured using conventional ac
four-terminal lock-in methods, whereas TDUCF and MFUCF were measured
using an ac five-terminal bridge technique\cite{Scofield87RSI}.
Excitations were restricted to levels that avoided Joule self-heating
of the carriers, checked by comparing temperature dependences of
measurements at different drive amplitudes.  This requirement sets the
lower temperature limit of the noise measurements.

Typical In$_{1-x}$Mn$_{x}$As is paramagnetic above its Curie
temperature, $T_{\rm c}$, which is between 10 and 100~K depending on
Mn concentration, substrate temperature, and resulting disorder during
the growth process\cite{SchallenbergetAl06APL}.
Fig.~\ref{fig:Resistance} (a) and (b) show the temperature dependence
of the longitudinal and Hall resistance, respectively, for sample \# 1
where as fig.~\ref{fig:Resistance} (c) and (d) are for sample \# 2.
The resistance peaks at $T_{\rm c}$ between 30-50~K while cooling, and
the transition is clear in the Hall resistance, which exhibits
ferromagnetic hysteresis for $T<T_{\rm c}$ (grey lines).  This is more
pronounced as $T$ is further reduced (black lines).  In the TDUCF and
MFUCF measurements, $T$ is well below $T_{\rm c}$.

\begin{figure}[h!]
\begin{center}
\includegraphics[clip, width=7.7cm]{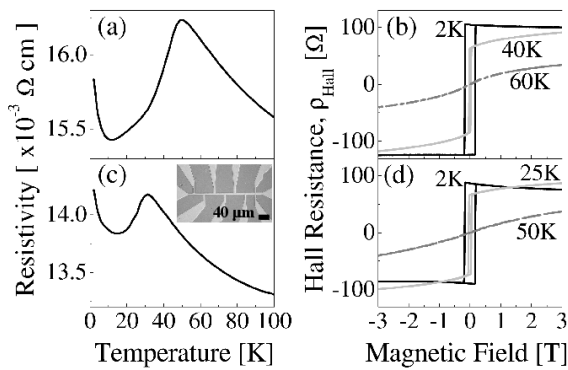}
\end{center}
\vspace{-3mm}
\caption{(a) Resistivity of sample \# 1 shows its Curie temperature
around 50~K. (b) Hall resistance measurement for sample \# 1 at 60
(dashed line), 40 (grey line), and 2~K (black line) showing magnetic
hysteresis upon cooling, as expected for the paramagnetic to
ferromagnetic transition. (c) Resistivity for sample \# 2 shows its
Curie temperature lies $\sim$ 30~K. (d) Hall resistance for sample \#
2 at 50 (dashed line), 25 (grey line) and 2~K (black line). Inset: SEM
image of etched sample.}
\label{fig:Resistance}
\vspace{0mm}
\end{figure}

Figure~\ref{fig:TDUCF} shows the results of the TDUCF noise power,
$S_R (T)$, measurements on the two samples.  The frequency dependence
of the raw voltage noise power\cite{supp} is well described as $1/f$.
The parameter plotted is the coefficient of the $1/f$ dependence,
normalized by drive current and sample resistance to units of 1/Hz.
The noise floor of the measurement setup has been subtracted from
these data.  Closed (open) symbols indicate the
perpendicular (parallel) field configuration.  The {\it increase} of
noise power as $T$ is {\it decreased} is a unique, distinguishing
feature expected in TDUCF, and results from the growth of the
coherence length, $L_{\phi}$\cite{LeeetAl04PRB}.

\begin{figure}[h!]
\begin{center}
\includegraphics[clip, width=7cm]{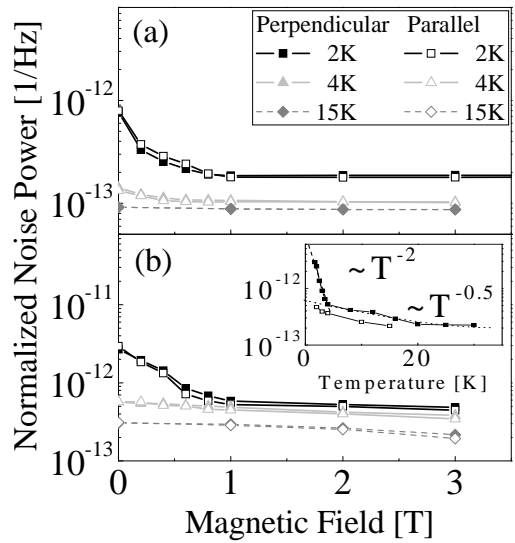}
\end{center}
\vspace{-5mm}
\caption{Normalized noise powers as a function of external magnetic 
field for (a) sample $\#$ 1, and (b) sample $\#$ 2 are shown at 
three different temperatures, 2~K (solid black lines), 4~K (solid 
light grey lines), and 15~K (dashed grey lines). Closed symbols are 
for perpendicular configuration, and open symbols are for parallel 
configuration. Inset shows the temperature dependence of noise power 
for sample $\#$ 2 at 0 T (solid symbol) and 3 T (open symbol). Error 
bars are not shown in these plots because they are comparable to the 
symbol size.}
\label{fig:TDUCF}
\vspace{-5mm}
\end{figure}

The inset to Fig.~\ref{fig:TDUCF}(b) shows $S_{R}(T)$ at zero-field
and $B = 3$~T for sample \#2; the same trends are seen in sample \#1.
Above $\sim$ 5~K at $B=0$ as well as over the whole temperature range
for the high field data, the noise power varies approximately as
$T^{-0.5}$.  In the low temperature limit, the noise power scales
approximately as $T^{-2}$.  Both of these dependences differ from the
temperature dependence seen in nonmagnetic metals.  In normal metals,
the expected\cite{FengetAl86PRL,BirgeetAl89PRL} $S_{R}(T) \sim
n(T)L_{\rm min}^{2}L_{\phi}^{2}$, where $n(T)$ is the density of
thermally active fluctuators, $L_{\rm min}$ is the smaller of
$L_{\phi}$ or $L_{T}\equiv \sqrt{\hbar D k_{\rm B}T}$, the thermal
length.  For standard two-level fluctuators, $n(T)\sim T$, and in
typical metals, $L_{T}<L_{\phi}$, implying that $S_{R}\sim
L_{\phi}^{2}$.  The unusual temperature dependence seen in FS samples
therefore suggest either (a) the fluctuators have an unusual energy
distribution despite having the usual distribution of relaxation times
that gives $S_{R}\sim 1/f$; or (b) the dephasing mechanism for holes
in this material is unconventional.

In normal metals, the field-dependent suppression of the cooperon
contribution to the TDUCF provides a means of quantitatively assessing
$L_{\phi}$ without recourse to these assumptions about the
distribution of fluctuators\cite{Stone89PRB}.  In a ferromagnetic
system with broken time-reversal symmetry, {\it no such decrease} in
noise power is expected, since the cooperon contribution is likely
already suppressed.  This was borne out in permalloy
wires\cite{LeeetAl04PRB}.  In contrast, $S_{R}(B)$ in the
In$_{1-x}$Mn$_{x}$As system is striking at low temperatures, and
strongly implies that the dominant source of TDUCF is coherent
scattering of carriers off fluctuating magnetic disorder.
Qualitatively similar effects have been seen at milliKelvin
temperatures in magnetic semiconductors that are spin
glasses\cite{JaroszynskietAl95PRL}.  Above 5~K, $S_{R}$ is essentially
independent of $B$, as seen in the permalloy
experiments\cite{LeeetAl04PRB}.  However, as $T$ is reduced below 5~K,
$S_{R}$ acquires a field dependence for $|B|< \sim$~1~T.
For both samples, $S_R (B~=~0~T)$ becomes almost four times larger
than $S_R (B~=~1~T)$ at 2~K.  As shown in Fig.~\ref{fig:TDUCF}, this
field dependence is approximately independent of field orientation.
This is consistent with a field-driven Zeeman suppression of the
fluctuators that cause the noise, rather than an orbital coherence
effect as in TDUCF in normal metals.

UCF as a function of magnetic field (MFUCF) rather than time provide a
consistency check on the idea that Zeeman rather than orbital physics
is relevant to $S_{R}(B)$.  Since Aharonov-Bohm shifting of phases of
electronic trajectories is equivalent to altering the impurity
configuration\cite{LeeetAl87PRB}, sweeping $B$ leads to
sample-specific, reproducible MFUCF within a coherent volume.  The
correlation field scale of the fluctuations, $B_{\rm c}$, is related
to the size of typical coherent trajectories via the flux quantum,
$h/e$.  Figure~\ref{fig:MFP} (a) shows MFUCF for sample \# 1 in three
different temperatures 2, 4, and 10~K from $B$= -9~T to 9~T when
sample lies in perpendicular configuration.  The symmetry in $B$ about
$B=0$ outside of the hysteretic portion of the conduction response
confirms that these fluctuations are real.  Qualitatively similar
MFUCF occur in sample \#2.  Note that the MFUCF reproduce when 
$B$ is swept up and back.  We assume a quasi-2d response, $B_{\rm
c}L_{\phi}^{2}\sim h/e$, and check for consistency.  While far more
fluctuations are required for a firm quantitative estimate,
Fig.~\ref{fig:MFP} (b) suggests that $B_{\rm c}$ at 2~K is on the
order of 2~T, which would imply coherence length, $L_{\phi} \sim$
50~nm, larger than sample thickness as required for self-consistency.
A truly quasi-2d sample would exhibit weaker fluctuations in the
parallel configuration, and this is consistent with Figs.~\ref{fig:MFP}(c)
and (d).  The MFUCF variance varied like $\sim T^{-4}$ dependence
below 4~K, stronger than the TDUCF variance, though statistics are
poor because the apparent field scale of the MFUCF is so large.

\begin{figure}[h!]
\begin{center}
\includegraphics[clip, width=7.5cm]{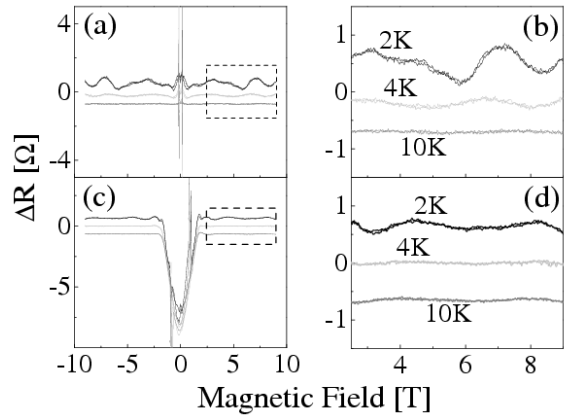}
\end{center}
\vspace{-3mm}
\caption{MFUCF measurements of $\Delta R \equiv R(B)-R(B=0)$ for sample \# 1 in (a) perpendicular 
and (c) parallel configurations were done using the 
five-terminal bridge technique.  Measurements were performed at 
three different temperatures 2, 4, and 10~K for the field span 
of $-$9~T to +9~T.  The curves are offset for clarity and 
a smooth background magnetoresistance due to imperfect 
symmetry between the two sides of the bridge is subtracted 
from each curve.  (b) and (d) are zoomed-in version for the 
marked area in (a) and (c), respectively. }
\label{fig:MFP}
\vspace{0mm}
\end{figure}

The microscopic origins of the TDUCF remain a subject for further
investigation.  One possibility is that the fluctuators are associated
with Mn spins, perhaps at the edges of the sample, not fully
participating in the bulk FM order of the system.  Coherent scattering
off slowly fluctuating local magnetization could cause TDUCF, as in
the spin glass case mentioned previously\cite{JaroszynskietAl95PRL}.
At sufficiently large $B$ those moments would be saturated, removing
that source of fluctuations.  This does not, however, explain the
particular forms of $S_{R}(B, T)$, and the similarity in temperature
dependence in the low $T$ limit to that seen previously in permalloy
wires.

In summary, we have examined quantum corrections to the electronic
conduction in InMnAs nanostructures.  The observed dependences of
TDUCF and MFUCF on temperature and magnetic field place constraints on
the noise and decoherence processes at work in this material.  InMnAs
is another rich laboratory in which to examine quantum coherence in
the presence of ferromagnetism, physics that may enable future device
technologies.

% \vspace*{3mm}

This work was supported by DOE grant DE-FG03-01ER45946/A001 and the
David and Lucille Packard Foundation.  TS and HM acknowledge support
in part by Grant- in-Aid for Scientific Research from MEXT and JSPS
(No. 14076210 and No. 17206002), and by the NSF-IT program
(DMR-0325474) in collaboration with Rice University.

%\clearpage

%%%%%%%%%%%%%%%%%%%%%%%%%%

%\clearpage

%\clearpage

\clearpage

\title{Supplementary material for ``Mesoscopic conductance effects in InMnAs structures''}

\author{S. Lee$^{1}$, A. Trionfi$^{1}$, T. Schallenberg$^{2}$, H. Munekata$^{2}$, and D. Natelson$^{1}$}
\affiliation{$^{1}$Department of Physics and Astronomy, Rice University, Houston, TX 77005, USA}
\affiliation{$^{2}$Imaging Science and Engineering Laboratory, Tokyo Institute of Technology, 
Yokohama, Kanagawa 226-8503, Japan}

\date{\today}

\maketitle

Due to space limitations in the main manuscript, here we show
two supplemental figures that provide further information about
the experiments.

\begin{figure}[h!]
\begin{center}
\includegraphics[clip, width=8 cm]{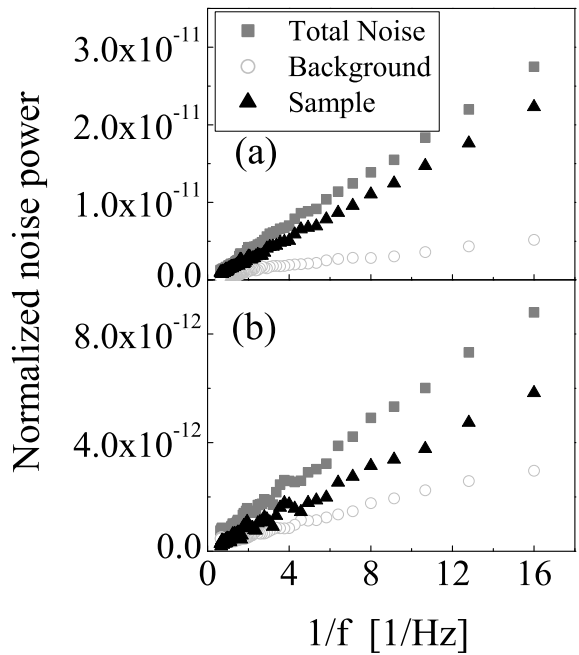}
\end{center}
\caption{Normalized raw noise data for sample \# 1, for (a) $T = 2$~K, $B = 0$~T, and
(b) $T = 15$~K, $B = 3$~T, with the field oriented perpendicular
to the plane of the sample.} \label{fig:supfig1}
\end{figure}

Figure~\ref{fig:supfig1} shows the measured noise power from the
in-phase channel of the lock-in amplifier (squares), the measured
system background noise from the out-of-phase channel of the lock-in (circles),
and their difference (triangles), for the high noise (a) and low
noise (b) limits of the experiment, plotted versus $1/f$.  The 
noise signals in $V^{2}$ corrected for amplifier gain 
have been normalized by $(I_{\rm dr}R)^{2}$, where $I_{\rm dr}$ is
the drive current used for the measurement.  Note that even
at 15~K and 3~T, the noise signal is readily detectable above
the background noise of the circuit.

\begin{figure}[h!]
\begin{center}
\includegraphics[clip, width= 8 cm]{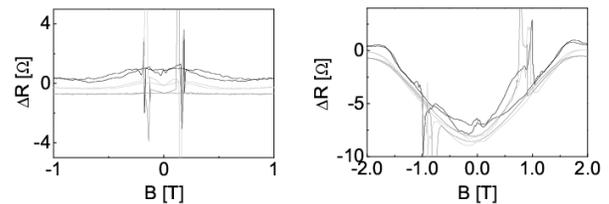}
\end{center}
\caption{Rescaled plots of the data from Fig.~3 of the main paper, 
showing the hysteretic region of the magnetofingerprint data for
(left) $B$ perpendicular to the sample plane, and (right) $B$ along
the direction of the sample.  Top to bottom at high fields, the data
are 2~K, 4~K, and 10~K.}
\label{fig:supfig2}
\end{figure}

Figure~\ref{fig:supfig2} shows rescaled plots of the low-field regions
of the magnetofingerprint data from Fig.~3 of the paper, for
both field orientations.  Note that these data were taken using
the same five-terminal bridge method as the noise data.  In an
ideal sample configuration, the two sides of the bridge would 
be exactly symmetrical, and bulk magnetoresistive effects such
as the anisotropic magnetoresistance would be perfectly nulled
away.  The low-field regions shown here demonstrate that the 
actual samples are slightly unsymmetric, such that the field-driven
reorientation of the sample magnetization does not happen 
uniformly across the whole sample.  As a result, there is some
hysteresis seen in these plots as the magnetization is coerced
at slightly different fields in the two halves of the sample.  
In the parallel field configuration, the main effect seen is
the un-nulled portion of the anisotropic magnetoresistance,
as $M$ is coerced away from the perpendicular direction favored
by crystallographic anisotropy, and into the direction of current
flow.

\end{document}